\documentclass[prl,superscriptaddress,twocolumn,amsmath,amssymb]{revtex4}

\usepackage{graphicx}
\usepackage{bm}
\usepackage{ulem}

\usepackage{color} 

\definecolor{Red}{rgb}{1,0,0}
\definecolor{Blue}{rgb}{0,0,1}

\usepackage{lscape}
\newcommand{\Tonee}{\ensuremath{T_{1\mathrm{e}}}}
\newcommand{\mi}{\ensuremath{m_{\mathrm{I}}}}

\newcommand{\figurebox}[1]{#1}

\newcommand{\replace}[2]{#2}
\newcommand{\strikeout}[1]{}

\newcommand {\Fig}[1] {Figure~\ref{#1}}

\newcommand{\beq}{\begin{equation}}
\newcommand{\eeq}{\end{equation}}

\newcommand{\pthirtyone}{$^{31}$P}

\newcommand{\sitwonine}{$^{29}$Si}
\newcommand{\biiso}{$^{209}$Bi}

\newcommand{\beqa}{\begin{eqnarray}}
\newcommand{\eeqa}{\end{eqnarray}}

\newcommand{\JMR}{J. Mag. Res.}

\newcommand{\PR}{Phys. Rev.}

\newcommand{\PRB}{Phys. Rev. B}
\newcommand{\PRL}{Phys. Rev. Lett.}

\newcommand{\ttwo}{$T_2$}

\newcommand{\tone}{$T_1$}

\newcommand{\tonee}{$T_{\rm{1e}}$}
\newcommand{\ttwon}{$T_{\rm{2n}}$}

\newcommand{\tsd}{$T_{\rm SD}$}

\newcommand{\natsi}{$^{\rm nat}$Si}
\begin{document}

\title{Electron spin coherence and electron nuclear double resonance of \\ Bi donors in natural Si}

\author{Richard E. George}
\affiliation{CAESR, Clarendon Laboratory,Department of Physics, Oxford University, Oxford OX1 3PU, UK}

\author{Wayne Witzel}
\affiliation{Sandia National Laboratories, Albuquerque, New Mexico 87185, USA}

\author{H. Riemann}
\author{N.V. Abrosimov}
\author{N. N\"{o}tzel}
\affiliation{Institute for Crystal Growth, Max-Born Str 2, D-12489 Berlin, Germany}

\author{Mike L. W. Thewalt}
\affiliation{Dept. of Physics, Simon Fraser University, Burnaby, BC, Canada}

\author{John~J.~L.~Morton}
\email{john.morton@materials.ox.ac.uk}
\affiliation{CAESR, Clarendon Laboratory,Department of Physics, Oxford University, Oxford OX1 3PU, UK}
\affiliation{Department of Materials, Oxford University, Oxford OX1 3PH, UK}

\date{\today}

\begin{abstract}
Donors in silicon hold considerable promise for emerging quantum technologies, due to the their uniquely long electron spin coherence times. Bismuth donors in silicon differ from more widely studied Group V donors, such as phosphorous, in several significant respects: they have the strongest binding energy (70.98~meV), a large nuclear spin ($I=9/2$) and strong hyperfine coupling constant ($A=1475.4$~MHz). 
These larger energy scales allow us to perform a detailed test of theoretical models describing the spectral diffusion mechanism that is known to govern the electron spin decoherence of P-donors in natural silicon. We report the electron nuclear double resonance spectra of the Bi donor, across the range 200~MHz to 1.4~GHz, and confirm that coherence transfer is possible between electron and nuclear spin degrees of freedom at these higher frequencies.
\end{abstract}
 

\maketitle

Electron and nuclear spin coherence of donors in silicon is of great importance for a number of proposals for Si-based quantum technologies~\cite{kane98, skinner03, vrijen00,morton10}. These schemes cite among the advantages for Si donor quantum bits:  long coherence times (exceeding tens of milliseconds for the electron and seconds for the nucleus in the case of Si:P), high-fidelity manipulation through a combination of microwave and radiofrequency pulses~\cite{mortonbb1}, and integration within silicon devices for measurement~\cite{morello09, tan10, hoehne10, lo07}. Magnetic resonance studies on P-donors in Si have examined electron spin coherence in natural Si~\cite{alexeisi,tyryshkin06} and its dependence on increasing \sitwonine~concentration in the host~\cite{abe04}, as well as the storage of coherent electron spin states in the \pthirtyone~nuclear spin~\cite{qmemory}. 

Although there has been a focus on the P-donor in Si, other Group V donors such as Bi also possess attractive qualities as quantum bits~\cite{stoneham03}. There have been relatively few recent studies on the Si:Bi, an exception being a \strikeout{recent} photoluminescence study showing dynamic nuclear \replace{polarisation}{polarization} of the \biiso~through optical pumping~\cite{thewalt10}. Bismuth is the deepest group V donor, with a binding energy of 70~meV~\cite{butler75} and the largest Group V hyperfine coupling of 1.4754~GHz~\cite{feher59} to the $I=9/2$ nuclear spin of \biiso. These parameters differ substantially from the P donor (44~meV and 117.52~MHz), raising the question of whether the same decoherence mechanisms and methodology for nuclear spin manipulation are applicable. In this letter we examine spin decoherence of Bi donors in natural silicon, as well as electron-nuclear double resonance (ENDOR) spectroscopy to probe the transitions of the \biiso~nuclear spin. We find that these measurements compare well with P donors in natural silicon, providing strong motivation to pursue $^{28}$Si:Bi material.

\begin{figure}[t] \centerline
{\includegraphics[width=3.5in]{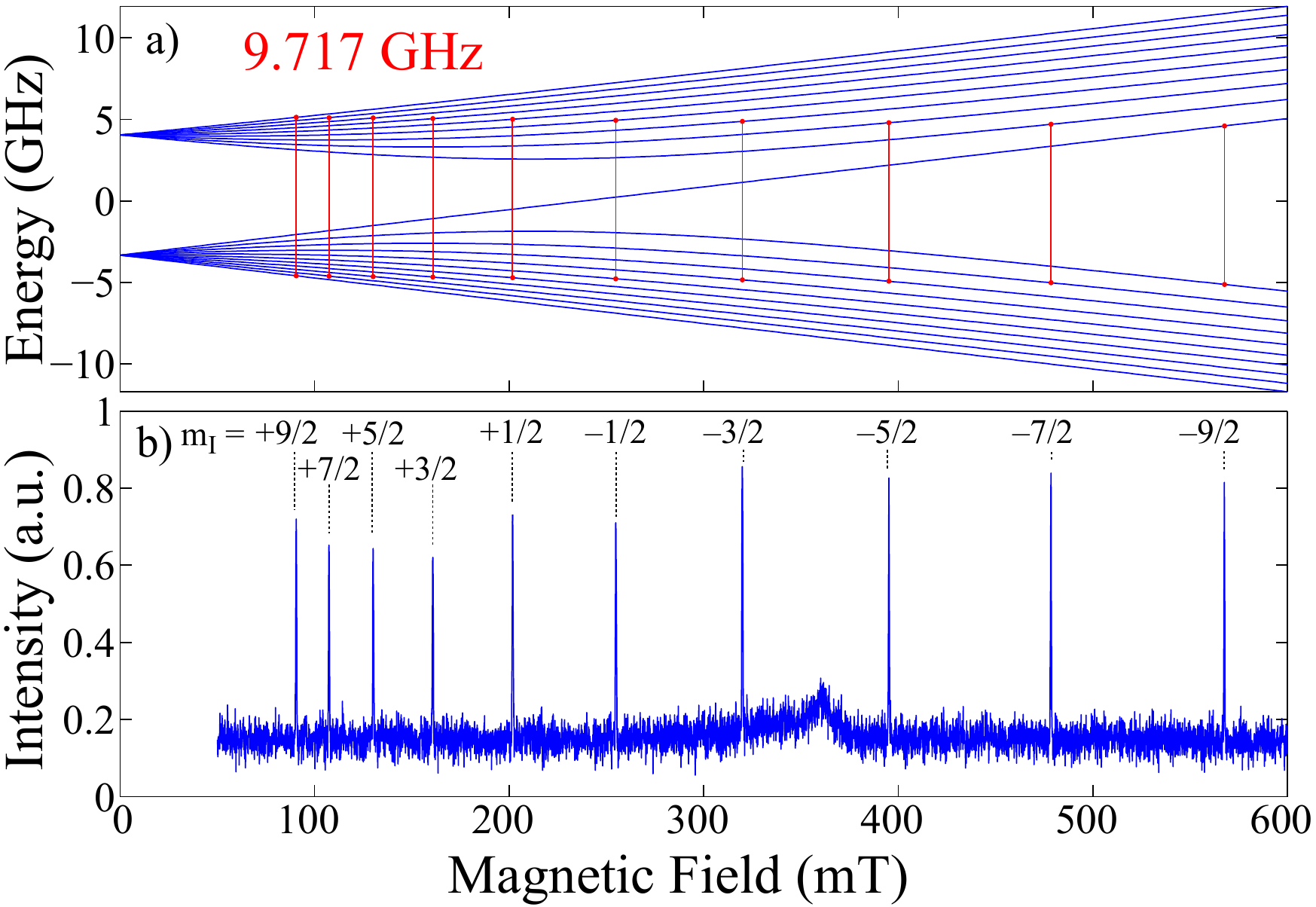}} \caption{(Colour online) Electron spin resonance of Si:Bi donors at X-band (9.7~GHz). a)~Energy levels of the coupled electron-nuclear spin system as a function of magnetic field. The allowed EPR transitions at 9.7~GHz are indicated with vertical lines, each corresponding to a different $m_I$ projection for the $I=9/2$ $^{209}$Bi nuclear spin. b) Experimental electron spin echo intensity as a function of magnetic field yields the 10 expected resonances. Temperature=16~K, simulation parameters: $A=1475.4$~MHz, $g_e=2.00$, $\gamma_{\rm Bi}=$6.962~MHz/T. 
} \label{ESEfswp}
\end{figure}

Natural Si:Bi samples were obtained from from ultrapure \natsi~starting material by a floating-zone technique, as described in Ref~\cite{riemann06} and had a room temperature resistivity of 4.5~$\Omega$cm implying a Bi concentration of $10^{16}$~cm$^{-3}$. Pulsed EPR measurements were performed using a Bruker Elexsys 680 X-band spectrometer, equipped with a low temperature helium-flow cryostat (Oxford CF935). A TWT amplifier was used to provide EPR $\pi/2$ and $\pi$ pulses of 20 and 40~ns respectively. For ENDOR measurements an Amplifier Research 60~W solid state CW amplifier (0.8--4.2~GHz) and an ENI 100~W (1.5--400~MHz) were used, depending on the frequency range, driven by an Agilent PSG Analogue Signal Generator. RF pulses in the range 4--15~$\mu$s were used for a $\pi$-pulse on the \biiso\ nuclear spin, depending on frequency. EPR/ENDOR spectra were simulated using Easyspin~\cite{stoll06}.

\begin{figure}[t] \centerline
{\includegraphics[width=3.5in]{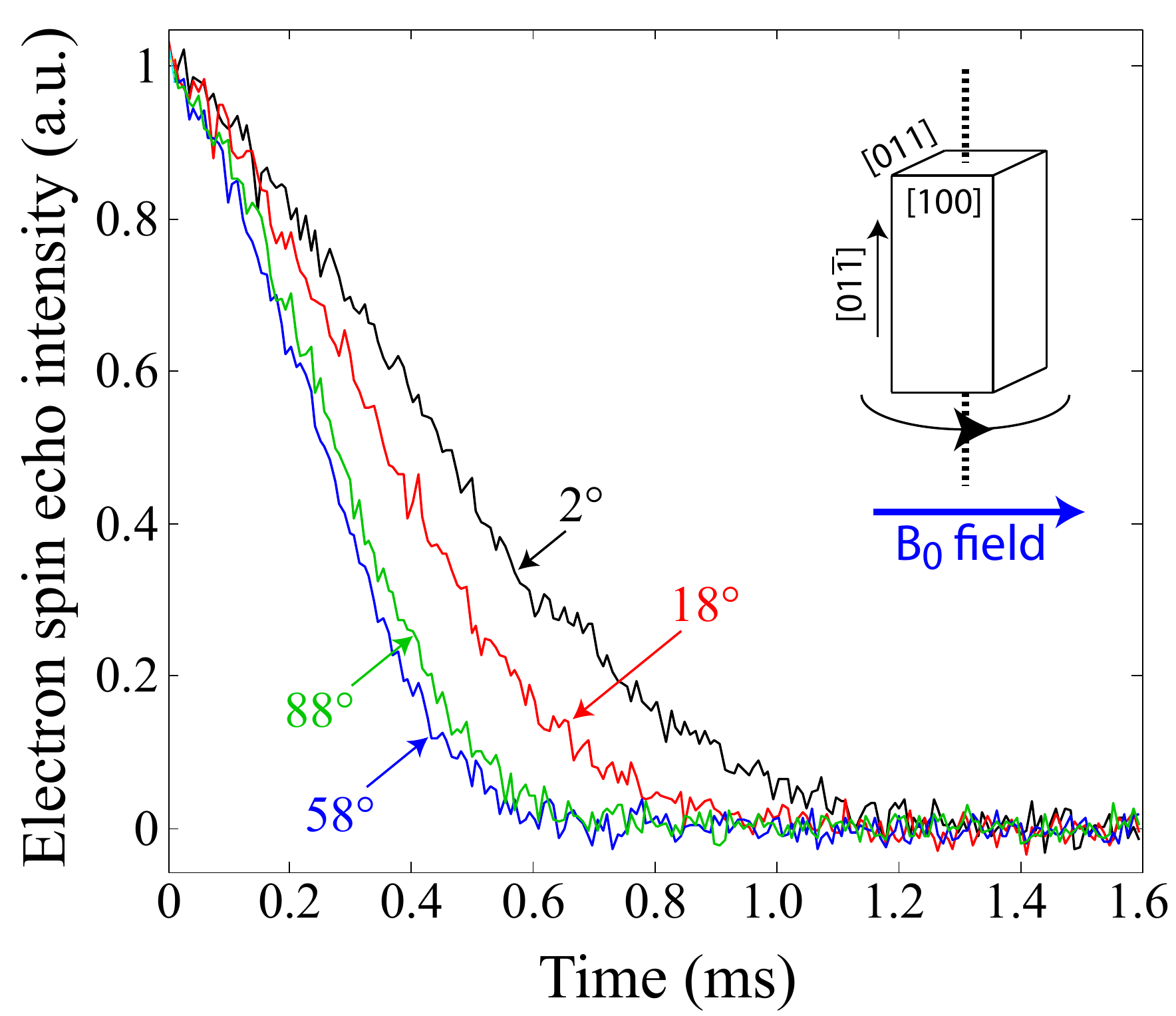}} \caption{(Colour online) Two-pulse electron spin echo decay of the Si:Bi donor as a function of angle of the applied magnetic field $B_0$ with respect to the [100] crystal axis. Crystal rotation is performed in the [100]--[011] plane. $T$ = 12~K, $B_0$ = 5663~G.} \label{t2sd}
\end{figure}

The Si:Bi electron/nuclear spin system can be described by an isotropic spin Hamiltonian (in angular frequency units):
\begin{equation}\label{Hamiltonian}
\mathcal{H}_0=\omega_e S_z - \omega_I I_z + A \!\cdot\! \vec{S} \!\cdot\!
\vec{I},
\end{equation}
where $\omega_e=g\beta B_0/\hbar$ and $\omega_I=g_I\beta_n B_0/\hbar$ are the electron and nuclear Zeeman frequencies, $g$ and $g_I$ are the electron and nuclear g-factors, $\beta$ and $\beta_n$ are the Bohr and nuclear magnetons, $\hbar$ is Planck's constant and $B_0$ is the magnetic field applied along $z$-axis in the laboratory frame. The donor electron spin S=1/2 (g = 2.0003) is coupled to the nuclear spin $I=9/2$ of $^{209}$Bi through a hyperfine coupling $A=1475.4$~MHz~\cite{feher59}. At high magnetic fields (i.e. $\omega_e \gg A(I+1/2)$), this leads to ten equally spaced resonances in the EPR spectrum, each corresponding to a transition $\Delta m_S = \pm1$ for a given $m_I$ projection. As shown in \Fig{ESEfswp}, measurements made at X-band (9.7~GHz) are not entirely in this high-field limit. The EPR spectrum of Si:Bi was recorded by monitoring the electron spin echo (ESE) intensity as a function of magnetic field. A linewidth of $\sim4$~G was measured for each of the ten EPR lines, consistent with inhomogeneous broadening from unresolved hyperfine coupling to the surrounding \sitwonine~nuclear spins ($\sim5\%$ natural abundance).

Electron spin decoherence of P-donors in natural silicon is known to be dominated by spectral diffusion, a mechanism in which spin flip-flop of surrounding \sitwonine\ nuclei modulates the electron Zeeman energy through both contact and dipolar hyperfine coupling~\cite{klauder62, salikhov81}. The \sitwonine\ nuclei closest to the donor are a `frozen core' which do not flip-flop due to the strong spatial dependence of their coupling to the donor electron spin, causing detuning between adjacent \sitwonine\ nuclear spins and suppressing nuclear flip-flop transitions that are allowed in the bulk material. Similarly, those furtherest away are too weak to influence the donor electron. There is therefore an `active shell' in the region where the dipolar coupling between neighbouring \sitwonine\ spins is comparable to their coupling to the donor electron spin, which is responsible for spectral spectral diffusion. This mechanism has been predicted~\cite{witzel06, sousa03} and shown experimentally~\cite{tyryshkin06} to have an angular dependence corresponding to the dependence of the dipolar coupling between nearest-neighbour \sitwonine\ spins on the crystal orientation with respect to the applied magnetic field. \Fig{t2sd} shows the measured electron spin echo decay traces as a function of angle, measured at the high-field line (5663~G, $m_I=-9/2$). The behaviour is qualitatively similar to that of P-donors in \natsi, though decay times are approximately 30$\%$ longer. This can be primarily attributed to the greater binding energy of the Bi donor compared with P, which shrinks the effective Bohr radius of the Si:Bi donor in comparison to Si:P, reducing the size of the 'active region'. There is also a secondary effect resulting from the much stronger hyperfine coupling to the donor nuclear spin, as described below. 

\begin{figure}[t] \centerline
{\includegraphics[width=3.5in]{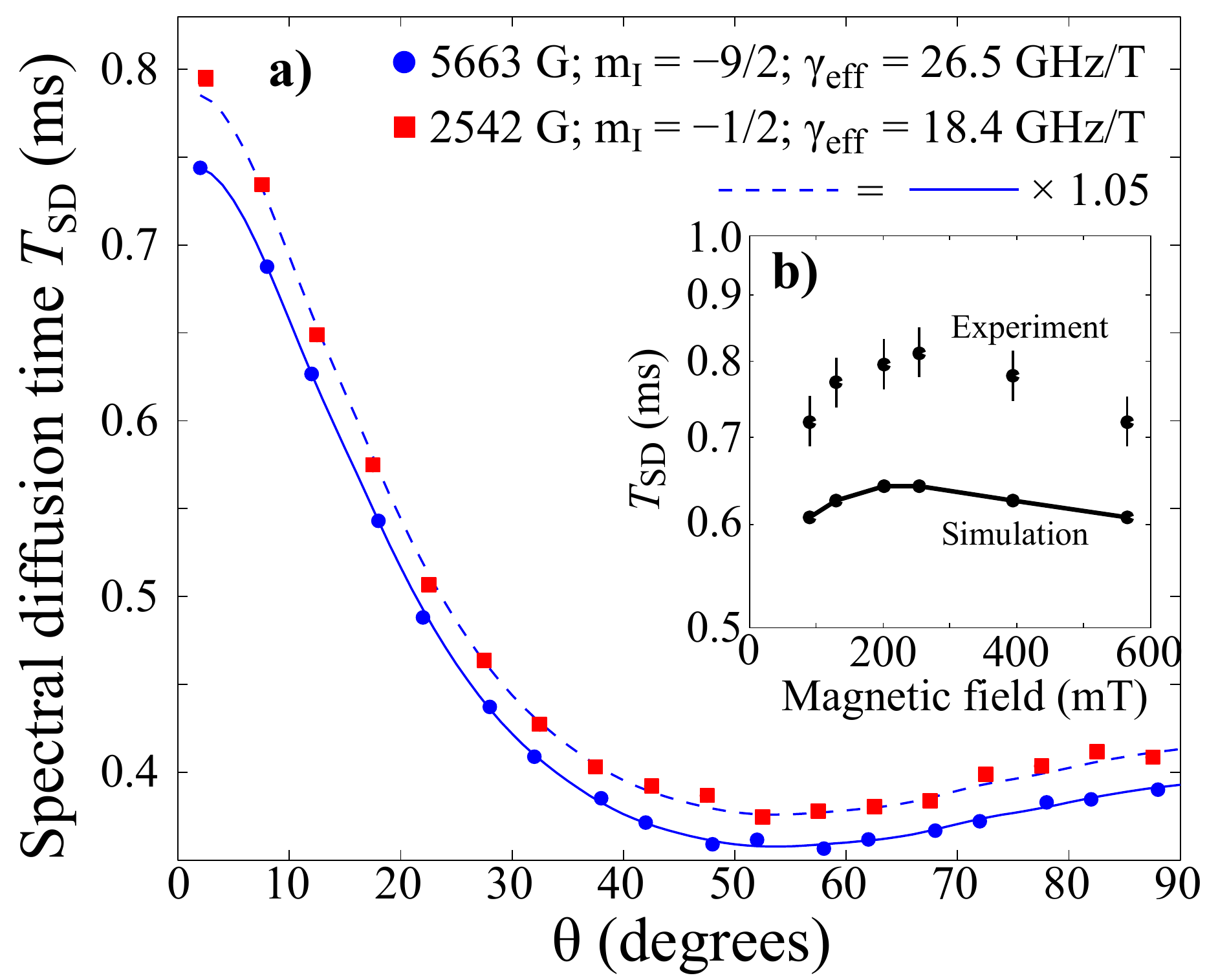}} \caption{(Colour online) Extracted spectral diffusion times, \tsd, as a function of crystal orientation and magnetic field. a) A fit to the electron spin echo decay curves provides a measure of \tsd~as function of angle of the applied magnetic field $B_0$ with respect to the [100] crystal axis, performed on two of the ten hyperfine lines: 5663~G (blue, circles) and 2542~G (red, squares). The solid curve is a spline fit through the data for 5663~G, which is multiplied by 1.05 to produce the dashed curve. b) Inset shows the magnetic field dependence of \tsd\ for six field positions ($\theta=0^{\circ}$), showing good agreement with the results of the simulation (see text). Temperature = 12~K.} \label{t2sdtheta}
\end {figure}

Although other Group V donors are well into the high-field approximation at X-band (9.7~GHz, 0.35~T), the large hyperfine coupling to \biiso\ causes some level mixing, as shown in \Fig{ESEfswp}. As a measure of the sensitivity of a transition frequency $f$ to a change in magnetic field $B$,  we can extract an effective gyromagnetic ratio $\gamma_{\rm eff} = df/dB$ which differs substantially from that of a free electron ($\gamma=28.0$~GHz/T), and also varies for each of the hyperfine lines across the EPR spectrum: from $\gamma_{\rm eff}\sim26.7$~GHz/T (for $m_I=\pm9/2$) to 18.4~GHz/T (for $m_I=\pm1/2$). The resulting change in $\gamma_{\rm eff}$ alters the coupling of the donor electron to surrounding \sitwonine\ and thus selects a different `active shell', with different statistics of pairwise \sitwonine\ coupling. Thus, though the dipolar coupling between neighbouring \sitwonine\ spins is primarily responsible for setting the timescale of spectral diffusion, we may expect an effect due to varying $\gamma_{\rm eff}$. Through simulations based on cluster expansion technique of Ref~\cite{witzel06} we have calculated the effect of varying $\gamma_{\rm eff}$, which predict a $\sim5\%$ increase in spectral diffusion times, \tsd, as measured on the $m_I=\pm1/2$ hyperfine line, compared to $m_I=\pm9/2$. The simulations use a Kohn-Luttinger wavefunction with the 70~meV binding energy for Bi.

We fit the echo decay traces, such as those in \Fig{t2sd}, to a combination of an orientation-independent \ttwo, combined with an orientation-dependent \tsd, through an expression of the form:
\beq
V(t)=V_0 e^{-(t/T_2)-(t/T_{SD})^{n}}
\eeq
The spectral diffusion mechanism has a characteristic stretched exponential coherence decay, with a typical value of $n$ between 1--4 depending on the regime of spectral diffusion~\cite{milov73, witzel06, sham06,loss08} . The stretching factor $n$ was found to be independent of crystal orientation, while \tsd, plotted in \Fig{t2sdtheta} shows the expected orientation dependence with a maximum when the applied magnetic field $B_0$ is oriented along [100], and a minimum when oriented along [111]. Comparing the values measured at two hyperfine lines: $m_I=-1/2$ (2542~G) and $m_I=-9/2$ (5663~G) we see \strikeout{excellent} agreement with the $\gamma_{\rm eff}$ dependence predicted by the simulations. The stretching factor $n$ showed no significant field dependence: $n_{\rm 2542G}=2.30(7)$ and $n_{\rm 5663G}=2.34(5)$, consistent with the simulations which predict 2.30(1).

The inset of \Fig{t2sdtheta} shows the predicted values of \tsd\ from the simulations, which are within $\sim20\%$ of the experimental values. The magnetic field dependence of \tsd\ is also well represented, showing a maximum for $m_I=\pm1/2$. The simulations were performed using the cluster correlation expansion~\cite{yang08} but it is sufficient to use a simple pair approximation that includes effects from each pair of nuclear spins independently~\cite{witzel05, sham06}.

The fitting error in the residual \ttwo~parameter is large when it is much longer than \tsd, however, our extracted values of a few milliseconds are consistent with being limited by electron spin relaxation (\tone) at this temperature (12~K).  
We find that \tone~is well described by a first-order Raman mechanism ($T^{-7}$) in the temperature range 8--16~K, as proposed by Castner for the range 19--25~K~\cite{castner63} (see Supporting Information).

\begin{figure}[t] \centerline
{\includegraphics[width=3.5in]{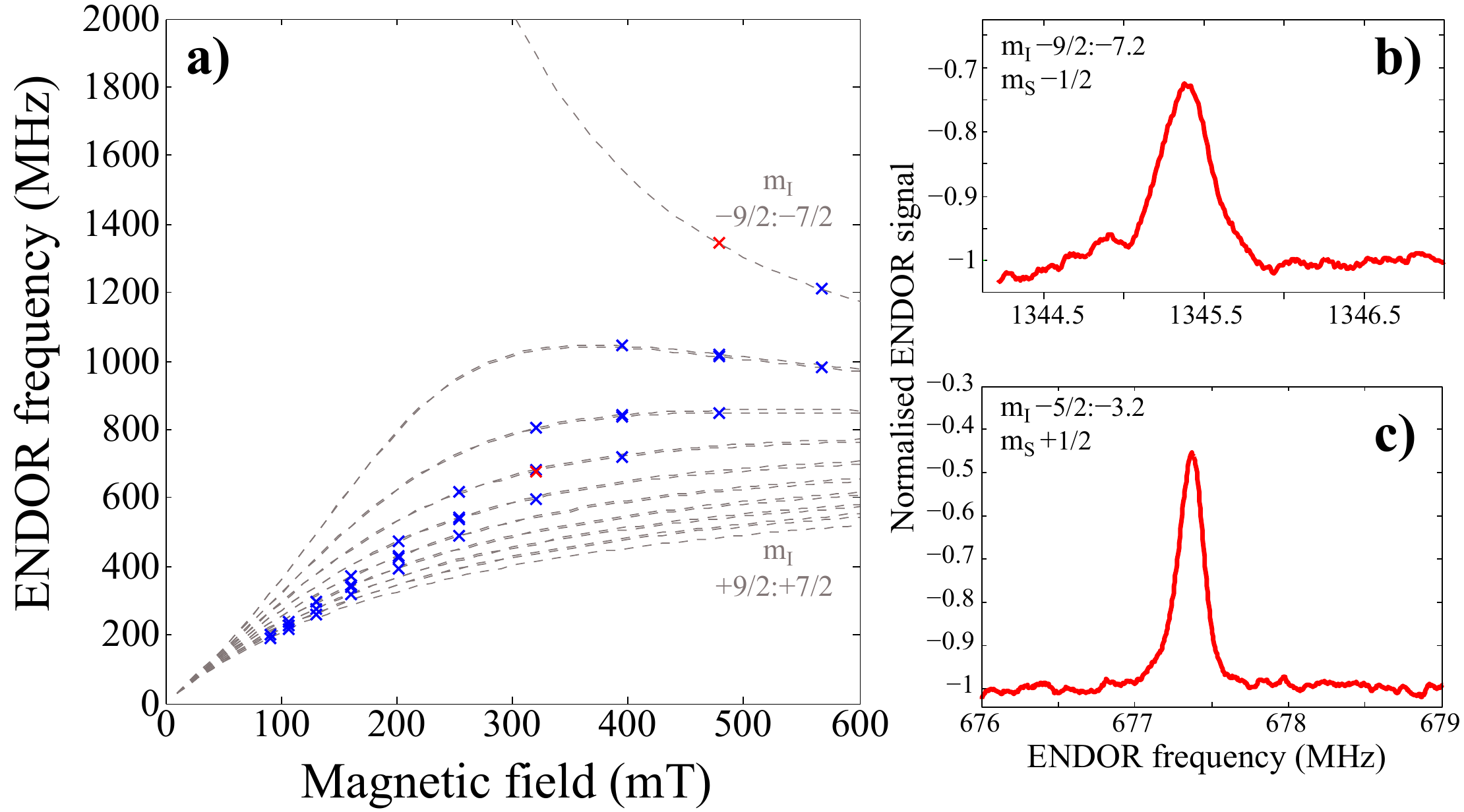}} \caption{(Colour online) Electron nuclear double resonance (ENDOR) of \biiso~in silicon. a) Dashed curves show the theoretical ENDOR frequencies as a function of field, for each $\Delta m_I=1$ transition. Symbols represent those values measured by Davies ENDOR at X-band, at each of the 10 resonant fields of the EPR spectrum. b) and c) show two typical ENDOR peaks, at 1345.4 and 677.4 MHz with RF $\pi$-pulse length = 4 and 7~$\mu$s, respectively.} \label{endorfig}
\end {figure}

We now turn to an investigation of the \biiso\ nuclear spin transition through electron-nuclear double resonance (ENDOR). We used the Davies ENDOR sequence ($\pi_{\rm mw} - \pi_{\rm rf} - \pi/2_{\rm mw} - \tau - \pi_{\rm mw} - \tau - {\rm echo}$) at each of the ten hyperfine lines to map out the set of 36 distinct ENDOR lines observable at a given microwave frequency: for each hyperfine line (i.e. EPR transition) there are four observable ENDOR transitions ($\Delta m_I =1$), apart from at the ends of the EPR spectrum ($m_I=\pm 9/2$) where there are only two. The measured frequencies are shown in \Fig{endorfig}a, along with theoretical curves showing ENDOR frequencies as a function of magnetic field. 

Two typical \biiso~ENDOR lines are shown in \Fig{endorfig}b and \ref{endorfig}c. The ENDOR linewidth ranges from 100--700~kHz, depending on the particular transition and field being measured. In general, the linewidths are broader at lower magnetic fields or rf frequencies, except for the two lines corresponding to the $m_I=-$9/2:$-$7/2 transition which are about 350~kHz wide.  
The ENDOR linewidths we observe (plotted in the Supporting Information) are well described by a combination of two factors: At lower magnetic fields ($< 400$~mT), the linewidth arises from the random dipolar field of surrounding \sitwonine, as in the case for the EPR linewidth. This effect is directly related to the gradient of the field/frequency curves shown in \Fig{endorfig}a which flatten out as the high-field approximation becomes valid (this also accounts for the greater linewidth of the $m_I=-$9/2:$-$7/2 transition). Correspondingly, this broadening mechanism is not  significant for ENDOR in other Group V donors at X-band given their much weaker hyperfine couplings. Instead, ENDOR linewidth in such donors arises from an inhomogeneity in the hyperfine coupling to the donor nucleus, due to a variation in the dielectric constant of the material within the donor wavefunction caused by the random distribution of \sitwonine. Such a mechanism could be responsible for the ENDOR linewidth in Si:Bi at higher magnetic field ($< 400$~mT) and from our measurements we can put an upper bound of the inhomogeneity in $a({\rm ^{209}Bi})$  to be ($< 0.02 \%$). 

It has been shown that the \pthirtyone~donor nuclear spin can provide a valuable resource for storing the coherent state of the electron spin for times exceeding seconds~\cite{qmemory}. The larger nuclear spin ($I=9/2$) of \biiso~provides a correspondingly larger Hilbert space for storing electron spin qubits, though also introduces more potential relaxation mechanisms for the nuclear spin. As we have found for Si:P, high-fidelity storage/retrieval of the electron coherence requires narrow EPR and ENDOR lines and thus a $^{28}$Si-enriched host material. We have investigated the potential of the \biiso~nuclear spin for quantum memory, and found that as expected, the \natsi~host limits the store/retrieve fidelity to $\sim$63$\%$ (see Supporting Information). We can nevertheless measure the decay of nuclear coherence and found that it is limited by the effect of \tonee\ (first order Raman) processes at temperatures above 10~K --- random fluctuations in the electron polarisation drive decoherence of the strongly coupled nuclear spin. Below this temperature, \ttwon\ is limited to about 15~ms. Further work using $^{28}$Si:Bi will be required to explore this limit and investigate the sources of nuclear decoherence.
 
We have found electron spin decoherence of Bi donors in natural Si to be dominated by the same spectral diffusion mechanism found in the case of Si:P, however thanks to the smaller Bohr radius of Bi the effect is weaker than for P, leading to 30$\%$ longer \ttwo~times. Despite the high ENDOR frequencies necessary to probe the \biiso~nuclear transitions, it is possible to excite each transition with a fidelity determined by the ENDOR linewidth of a few 100~kHz allowing us to demonstrate the possibility of storage and retrieval of electron spin coherence in the \biiso~nuclear spin. Applying such techniques to $^{28}$Si:Bi, we would anticipate the ability to store and retrieve multiple electron spin qubits with high fidelity within the nuclear spin. Finally, we note that the large energy splitting present at zero applied magnetic field makes the Si:Bi donor spin an attractive candidate for coupling to superconducting resonators. 

Note added: Interesting experimental and theoretical investigations of Si:Bi in the context of quantum information are reported in parallel studies~\cite{morley10,mohammady10}.

We thank Alexei Tyryshkin, Steve Lyon, Arzhang Ardavan and Andrew Briggs for helpful discussions. JJLM is supported by the Royal Society. The research is supported by the EPRSC through CAESR (No.\ EP/D048559/1). Sandia National Laboratories is a multi-program laboratory operated by Sandia Corporation, a wholly owned subsidiary of Lockheed Martin company, for the U.S. Department of Energy's National Nuclear Security Administration under contract DE-AC04-94AL85000.

\bibliography{biendor}
\clearpage

\section{Supplementary information}
\title{Supplementary information:\\ ``Electron spin coherence and electron nuclear double resonance of Bi donors in natural Si''}

\maketitle

 \makeatletter \renewcommand{\thefigure}{S\@arabic\c@figure} \renewcommand{\thetable}{S\@arabic\c@table} \makeatother 
 \setcounter{figure}{0} \setcounter{table}{0}

\subsection{Spin-lattice relaxation times (\Tonee)}

Electron spin-lattice relaxation times \Tonee\, were recorded as a function of temperature between 8 and 25~K using a three-pulse inversion recovery sequence $ \left( \pi \right)_x - \tau - \left( \pi/2 \right)_{\pm x} - \left( \pi \right)_x - \left( \mathrm{echo} \right) $ and two-step phase cycling.
 The inversion recovery plots were well fit to a mono-exponential decay and temperature was recorded with a calibrated thermometer below the microwave resonator. 

\begin{figure}[b]
\begin{center}
\figurebox{\includegraphics[width=3.3in]{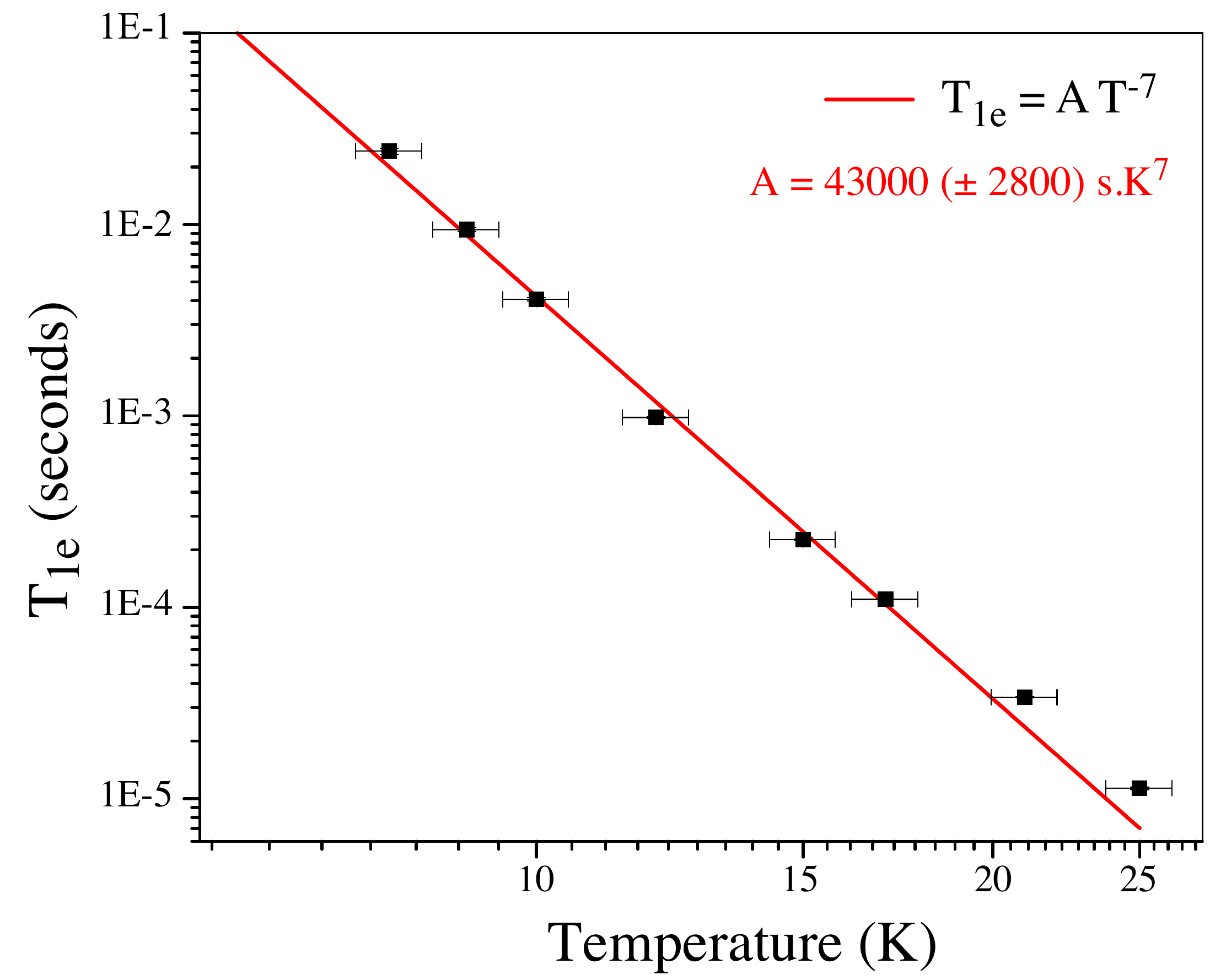}}
\caption{\label{raman_phonon_scattering_process}Temperature dependence of spin-lattice relaxation time in Si:Bi between 8 and 25K, showing a Raman-like $T^{-7}$ behaviour.}
\end{center}
\end{figure}

\Tonee\ is shown as a function of temperature in Fig. \ref{raman_phonon_scattering_process}, together with the least-squares fit to a model of the form $T_{1e} = A \, T^{-7}$, where $ A = 4.3 \left( \pm 0.3 \right) \times 10^4 \mathrm{ \, s \, K^{7}}$, consistent with the relaxation being driven by an inelastic Raman phonon scattering process \cite{castner}. 
The \Tonee\, data was collected from the $m_\mathrm{I} = -9/2$ line at 577.2~mT; we found \Tonee\,to be independent of the choice of \mi\,.

\subsection{Si:Bi ENDOR Linewidths}

\begin{figure*}
\begin{center}
\figurebox{\includegraphics[width=\textwidth]{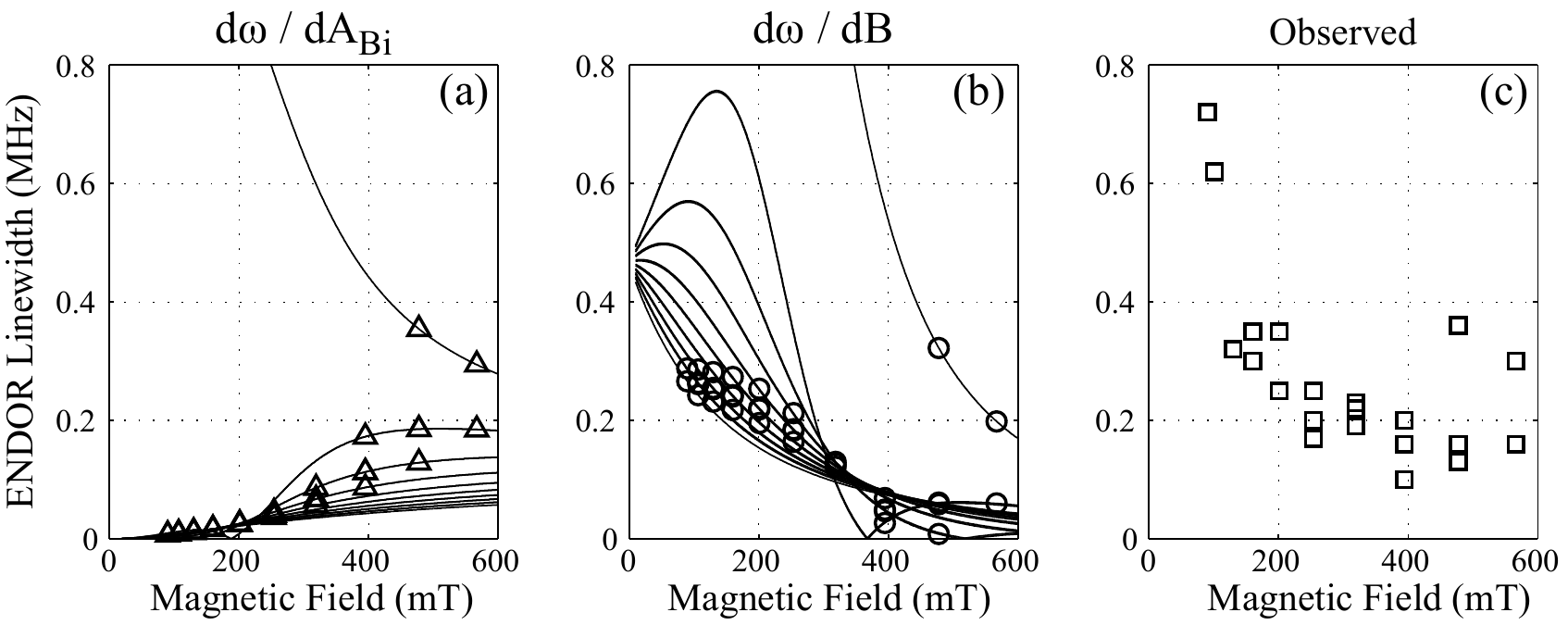}}
\caption{\label{ENDOR_linewidth_figure}ENDOR linewidths of hyperfine transitions in the Si:Bi systems observable using a microwave frequency of 9.7~GHz: \textbf{(a)} ($\triangle$) predicted due to a 0.23~MHz strain in the $^{209}$Bi hyperfine parameter $A$ (corresponding to 0.02$\%$), \textbf{(b)} ($\circ$) predicted due to random variation in magnetic field of $\sim2$~G. \textbf{(c)} ($\square$) Experimental data.}
\end{center}
\end{figure*}

Linewidths of the transitions between nuclear sublevels were recorded using a Davies ENDOR sequence: $\pi_{\rm mw} - \pi_{\rm rf} - \pi/2_{\rm mw} - \tau - \pi_{\rm mw} - \tau - {\rm echo}$. The RF pulse power was optimised by recording nuclear Rabi oscillations while the RF pulse length was chosen to avoid instrumental broadening of the ENDOR lines. ENDOR transition frequencies and linewidths were recorded on each of the ten EPR lines and the resulting distribution of linewidths is displayed in Fig. \ref{ENDOR_linewidth_figure}, together with theoretical calculations of the expected line broadening due to (a) strain in the hyperfine coupling to $^{209}$Bi and (b) random magnetic fields due to dipolar coupling to $^{29}$Si nuclear spins. The observed increase in linewidth at lower magnetic fields is consistent with the latter mechanism, though $A$-strain may contribute to ENDOR linewidth at fields above about 400~mT. 

\subsection{Fidelity of Si:Bi Nuclear Memory}

\begin{figure*}
\begin{center}
\figurebox{\includegraphics[width=\textwidth]{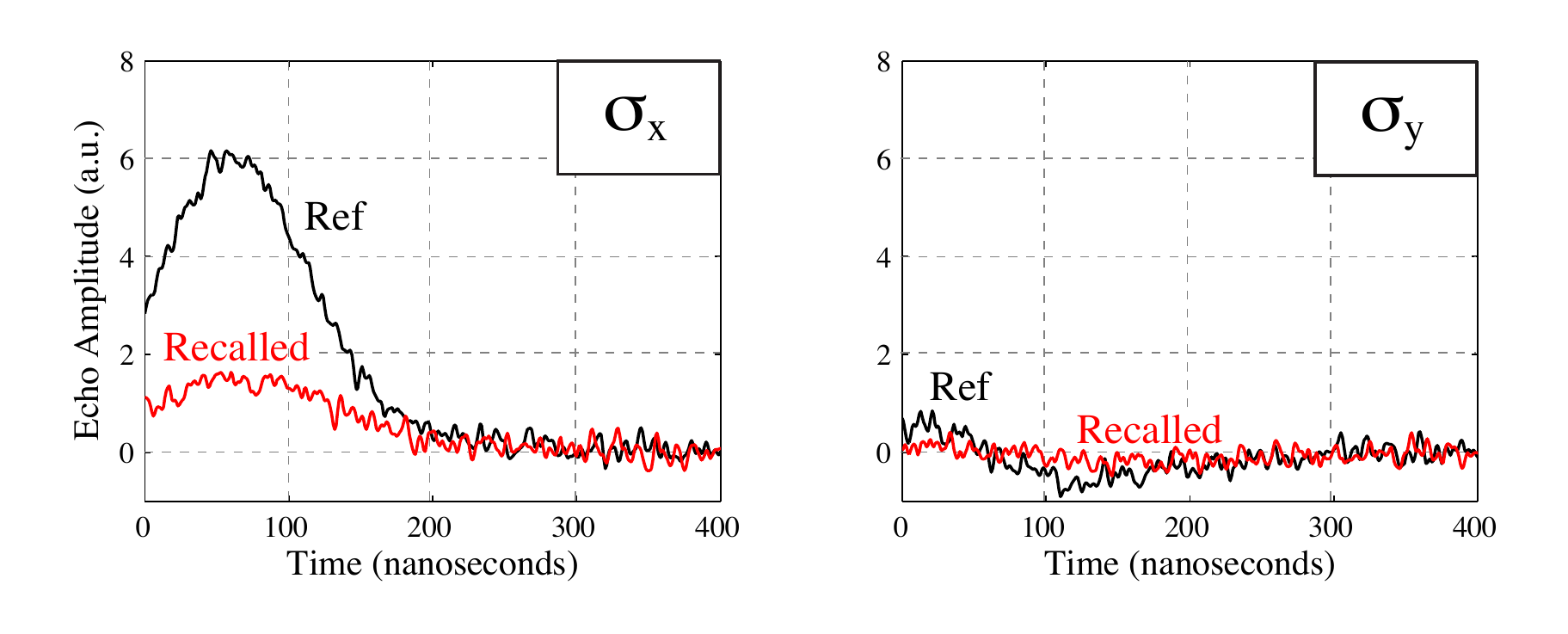}}
\caption{\label{Nuclear_memory_figure}Electron spin echo transients in Si:Bi comparing a reference (Ref) electron spin coherence with one that is stored in the nuclear spin and then recalled. The loss in signal intensity between the Reference and Recalled echoes is due to the imperfect fidelity of the coherence transfer process. For both cases, the real and imaginary components of the spin echo, measured using a quadrature detector, are shown separately. These signals are proportional to measurements of the $\sigma_x$ and $\sigma_y$ components of the density matrix, respectively~\cite{schweiger}.
Temperature = 14 K}
\label{nucmem}
\end{center}
\end{figure*}

Using the method described in Ref~\cite{Morton2008}, we demonstrate the ability to store some of the electron spin coherence in the nuclear spin, and subsequently recover it, as shown in Fig. \ref{nucmem}. It is not possible to fully excite the ENDOR hyperfine line, leading to a much lower transfer fidelity than observed for P-donors in isotopically purified $^{28}$Si, but comparable to that observed for $^{\rm nat}$Si:P.

Phase cycling of both microwave and rf pulses is used to confirm the recovered signal is a result of storage in the nuclear spin and is not an artefact such as a stimulated echo.

Comparing the intensity of the recovered electron spin echo with that which was stored allows for a rough assessment of the performance of the transfer --- in this case yielding $I_{\rm stored}\sim25\%$. 

To compare the transfer fidelity with other quantum memory experiments, we estimate the stored and recovered density matrices and compute the fidelity of that retrieved with respect to that stored using $\mathcal{F} =  {\rm Tr} \left( \rho_{\mathrm{ref}} \rho_{\mathrm{stored}} \right)$. We neglect the $\sigma_z$ component in both cases, and use the integrated electron spin echoes in the real and imaginary channels of the quadrature detector to extract $\sigma_x$ and $\sigma_y$ components. As in Ref~\cite{Morton2008}, the reference echo is designated as a pure state (i.e.\ no identity component) and used to normalise the intensity of the recovered echo. 

By this method, we estimate the fidelity of the store/retrieve process to be $\mathcal{F} = \left( 1 + I_{\mathrm{stored}} \right) / 2 = 0.635$. To avoid over-interpreting this result, it should be noted that recovering zero electron coherence (i.e.\ retrieving only the identity matrix) would yield a fidelity of 0.5.


\end{document}